\documentclass[a4paper]{jpconf}
\usepackage{graphicx}
\begin{document}
\title{Ultra High Energy Cosmic Rays an overview}

\author{Roberto Aloisio}

\address{Gran Sasso Science Institute, L'Aquila, Italy}
\address{INFN - Laboratori Nazionali Gran Sasso, Assergi (AQ), Italy}

\ead{roberto.aloisio@gssi.it}

\begin{abstract}
We review the main experimental evidences on ultra high energy cosmic rays and their implications in the physics of these extremely energetic particles, also in connection with dark matter and cosmology. We discuss the basis of theoretical models aiming at explaining observations, highlighting the most relevant open questions in this fascinating field of research. 
\end{abstract}

\section{Introduction}
Ultra High Energy Cosmic Rays (UHECR) are the most energetic particles ever observed, with detected energies that range from $10^{17}$ eV up to energies larger than $10^{20}$ eV.

The two running experiments devoted to the observation of UHECR are the Pierre Auger Observatory (Auger) in Argentina and Telescope Array (TA) in the US. UHECR can be observed only indirectly through the detection of their interaction products with the Earth's atmosphere. An UHE particle interacting with a nucleus of the atmosphere produces both hadronic and electromagnetic cascades of particles, collectively called Extensive Air Shower (EAS). In both Auger and TA the EAS detection is performed by directly detecting particles that reach the ground and by the observation of the fluorescence emission produced in the atmosphere by the EAS particles (due to excitation/de-excitation of atmosphere's Nitrogen molecules). 

Auger and TA, while sharing the same detection technique, based on the simultaneous observation of the EAS particles on ground and their fluorescence emission in air (hybrid events), exhibit a relevant difference in the acceptance, being that of Auger larger by roughly one order of magnitude.

As always in the case of cosmic rays, the information we can gather experimentally concerns: energy spectrum, mass composition and anisotropy. Since the first observation of an EAS with energy $~10^{20}$ eV back in 1962 \cite{Linsley:1963km}, the experimental study of UHECR clarified several important facts (see \cite{Coleman:2022abf,AlvesBatista:2019tlv} and references therein): (i) UHECR are charged particles, with a limit on photon and neutrino fluxes at $10^{19}$ eV at the level of few percent and well below respectively, (ii) the spectrum observed at the Earth shows a slight flattening at energies around $5\times 10^{18}$ eV (called the ankle) with (iii) an instep at $10^{19}$ eV and (iv) a steep suppression at the highest energies, (v) mass composition is dominated by light particles (proton and helium) at energies around $10^{18}$ eV becoming progressively heavier starting from energies around $3\times 10^{18}$ eV toward the highest energies.

In order to interpret the observations at the Earth, it is important a detailed modelling of UHECR propagation in the intergalactic medium, which is mainly conditioned by the interaction with astrophysical photon backgrounds and magnetic fields. These interactions shape the spectrum observed at the Earth and are responsible for the production of secondary (cosmogenic) particles: photons and neutrinos. This secondary radiation can be observed through ground-based or satellite experiments and contributes to the understanding of UHECR physics, bringing important information about the mass composition of UHECR and, possibly, on their sources. 

Sources of UHECR are still a mystery, we do not know which kind of astrophysical object is responsible for the production of these particles. There are basically two different classes of astrophysical mechanisms that could be invoked to accelerate UHECR \cite{Coleman:2022abf,AlvesBatista:2019tlv,Aloisio:2017ooo,Aloisio:2017qoo}. The first class is based on the transfer of energy from a macroscopic object (that can move relativistically or not) through repeated interactions of particles with magnetic inhomogeneities; belongs to this class the diffusive shock acceleration mechanism. The second class is based on the interaction with electric fields that, through high voltage drops, can accelerate particles (at  once) until the highest energies; belongs to this class the unipolar induction mechanism in fast spinning neutron stars \cite{Coleman:2022abf,AlvesBatista:2019tlv,Aloisio:2017ooo,Aloisio:2017qoo}. 

The extreme energies of UHECR, as high as $10^{11}$ GeV, eleven orders of magnitude above the proton mass and "only" eight below the Planck mass, are a unique workbench to probe new ideas, models and theories beyond the Standard Model (SM) of particle physics, which show their effects at energies much larger than those ever obtained, or obtainable in the future, in accelerator experiments. This is the case of theories with Lorentz invariance violations \cite{Adhikari:2022sve,Addazi:2021xuf,Aloisio:2000cm,Aloisio:2002ed} or models of Super-Heavy Dark Matter (SHDM) \cite{Adhikari:2022sve,PierreAuger:2022ibr,PierreAuger:2022wzk,Aloisio:2006yi,Guepin:2021ljb,Aloisio:2015lva,Aloisio:2007bh}, that connect UHECR observation with the dark sector and cosmological observations. 

The paper is organised as follows. In section \ref{sec:exp} it is given an overview of the current status of experimental measurements. In section \ref{sec:acc} are discussed the general requirements, in order to fit experimental evidences, of the astrophysical sources and acceleration mechanisms of UHECR. In section \ref{sec:shdm} it is discussed the impact of UHECR observations on models of super-heavy relic particles aimed at solving the Dark Matter (DM) problem. Conclusions take place in section \ref{sec:concl}.

\section{UHECR Observations} 
\label{sec:exp}

As discussed in the Introduction, nowadays the most advanced experiments observing UHECR are Auger and TA, in the present paper we will refer only to the results of these two experiments.

The high precision measurement of the spectrum of UHECR is of paramount importance to falsify theoretical models on the acceleration and propagation of UHECR. The integrated exposures reached so far by Auger and TA are respectively $12.2\times 10^4 $ km$^2$ sr y and $0.8\times 10^4$ km$^2$ sr y. Both experiments are based on the detection of hybrid events, with a set of fluorescence telescopes (FD) and a surface array of particles detectors (SD).

The determination of the UHECR energy in both Auger and TA is obtained by calibrating the fluorescence detection given the EAS fluorescence yield expected for a given energy of the primary particle. The ways Auger and TA implement this technique show a relevant difference. While Auger uses the absolute light yield and its wavelength behaviour as measured by the Airfly Collaboration (see \cite{Coleman:2022abf,AlvesBatista:2019tlv} and references therein), TA  uses the absolute florescence yield measured at fixed wavelength (337 nm) by Kakimoto et al. and the wavelength behaviour of the yield as determined by FLASH  (see \cite{Coleman:2022abf,AlvesBatista:2019tlv} and references therein). Another important difference between the two experimental techniques is related to the way the invisible energy (i.e. the portion of EAS energy not detectable) is treated. In the case of Auger this estimation is based on the Airfly measurements, while in the case of TA is obtained through MC simulations \cite{Coleman:2022abf,AlvesBatista:2019tlv}.

\begin{figure}[!h]
\centering
\includegraphics[scale=.555]{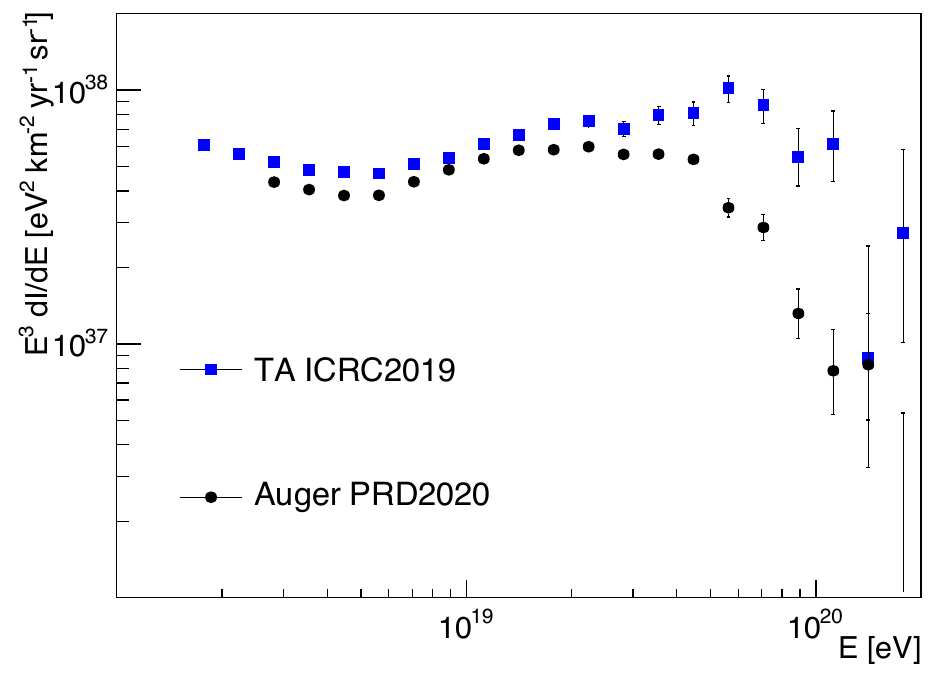}
\includegraphics[scale=.555]{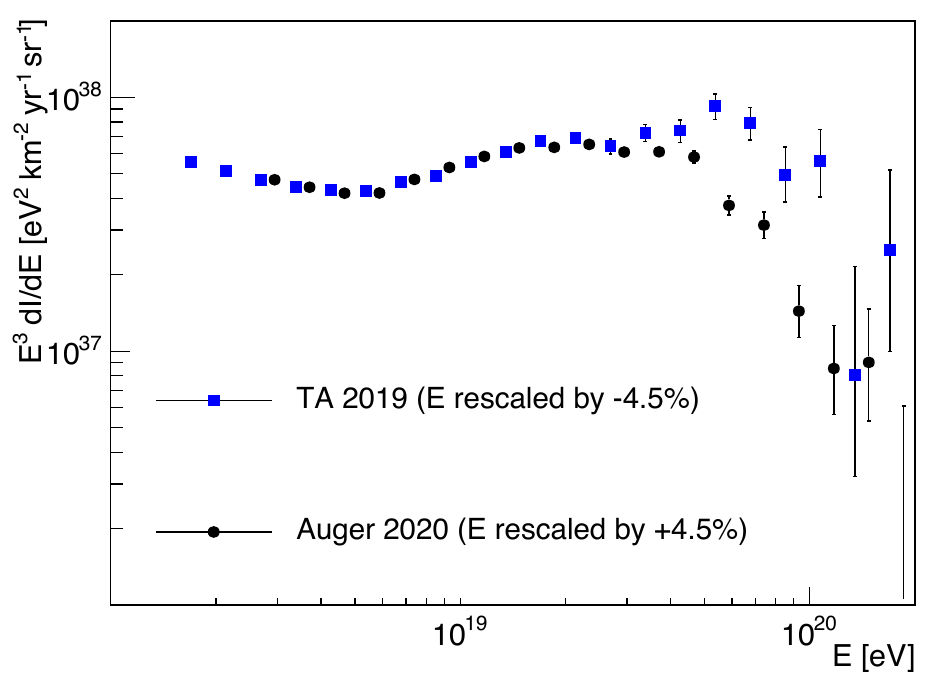}
\includegraphics[scale=.555]{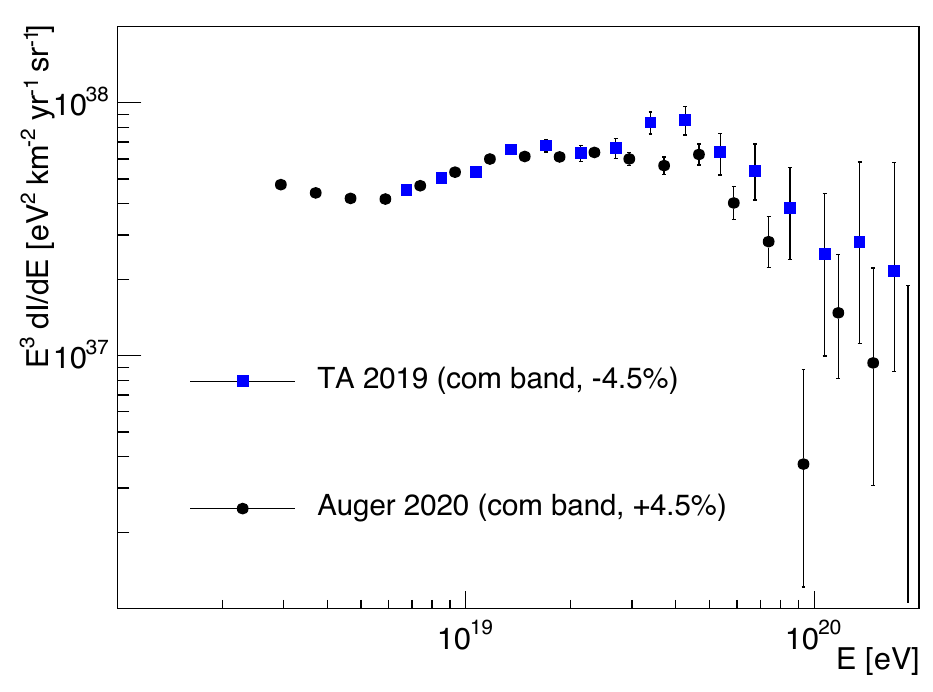}
\caption{In the left panel the comparison of the fluxes as measured by Auger and TA, in the central panel the fluxes with energy rescaled by +4.5\% (Auger) and -4.5\% (TA) and in the right panel the comparison of Auger and TA fluxes observed in the common declination band with energy rescaled as in central panel. Figure taken from \cite{Coleman:2022abf}.}
\label{fig1}  
\end{figure}

The joint working group on the spectrum measurement of the Auger and TA collaborations estimated the relative difference in the energy determination of the two experiments at the level of $6\%$ (see \cite{Coleman:2022abf,AlvesBatista:2019tlv} and references therein), that is perfectly compatible with the quoted systematic uncertainties of Auger $(14\%)$ and TA $(21\%)$. The UHECR fluxes measured by the two experiments are in a remarkable agreement up to energy around $3\times 10^{19}$ eV, see left panel of figure \ref{fig1}. Rescaling the Auger energies of $+4.5\%$  and the TA ones of $-4.5 \%$, well within the aforementioned uncertainties, the two observed fluxes show an excellent agreement apart from the highest energies, where the differences seem not conciliable (central panel of figure \ref{fig1}). To understand if this discrepancy has a physical origin related to the different sky portions observed by the two experiments, in the right panel of figure \ref{fig1} we have plotted the spectra observed using only UHECR events arriving in the common declination band of Auger and TA ($-15^\circ \le \delta \le 24.8^\circ$). Restricting the analysis to this sky portion the agreement improves, but at energies larger than $3\times 10^{19}$ eV the differences remain sizeable and cannot be explained by an independent scaling of the energy reconstruction in the two datasets. It should also be remarked that the energy spectrum measured by Auger does not show any dependence on the declination, while that observed by TA does, with some tension related to the expected UHECR anisotropy \cite{Coleman:2022abf,AlvesBatista:2019tlv}. Given the uncertainties in the energy determination and the poor statistics, it is too early to draw any definite conclusion on the reason of the differences between Auger and TA in the spectrum measurement at the highest energies.

Both experiments show a suppression in the energy spectrum at the highest energies. This effect is still not understood, as it can arise from UHECR propagation (through the processes of photo-pion production and photo-disintegration on astrophysical photons backgrounds) or it can be related to the maximum acceleration energy that the sources can provide \cite{Aloisio:2013hya,Aloisio:2015ega}.

The technique used by Auger and TA to determine the mass composition of UHECR is based on the measurement of the depth $X_{max}$ in the atmosphere at which the number of particles in the EAS reaches its maximum and the relative primary particle energy. These measurements are obtained by the FD observation. While in the case of Auger, thanks to enough statistics, the $X_{max}$ determination is obtained independently of detector effects, in the case of TA it is impossible, due to the reduced statistics, and the detector effects should be properly subtracted to compare with Auger results (see \cite{Coleman:2022abf,AlvesBatista:2019tlv} and references therein).

\begin{figure}[!h]
\centering
\includegraphics[scale=.55]{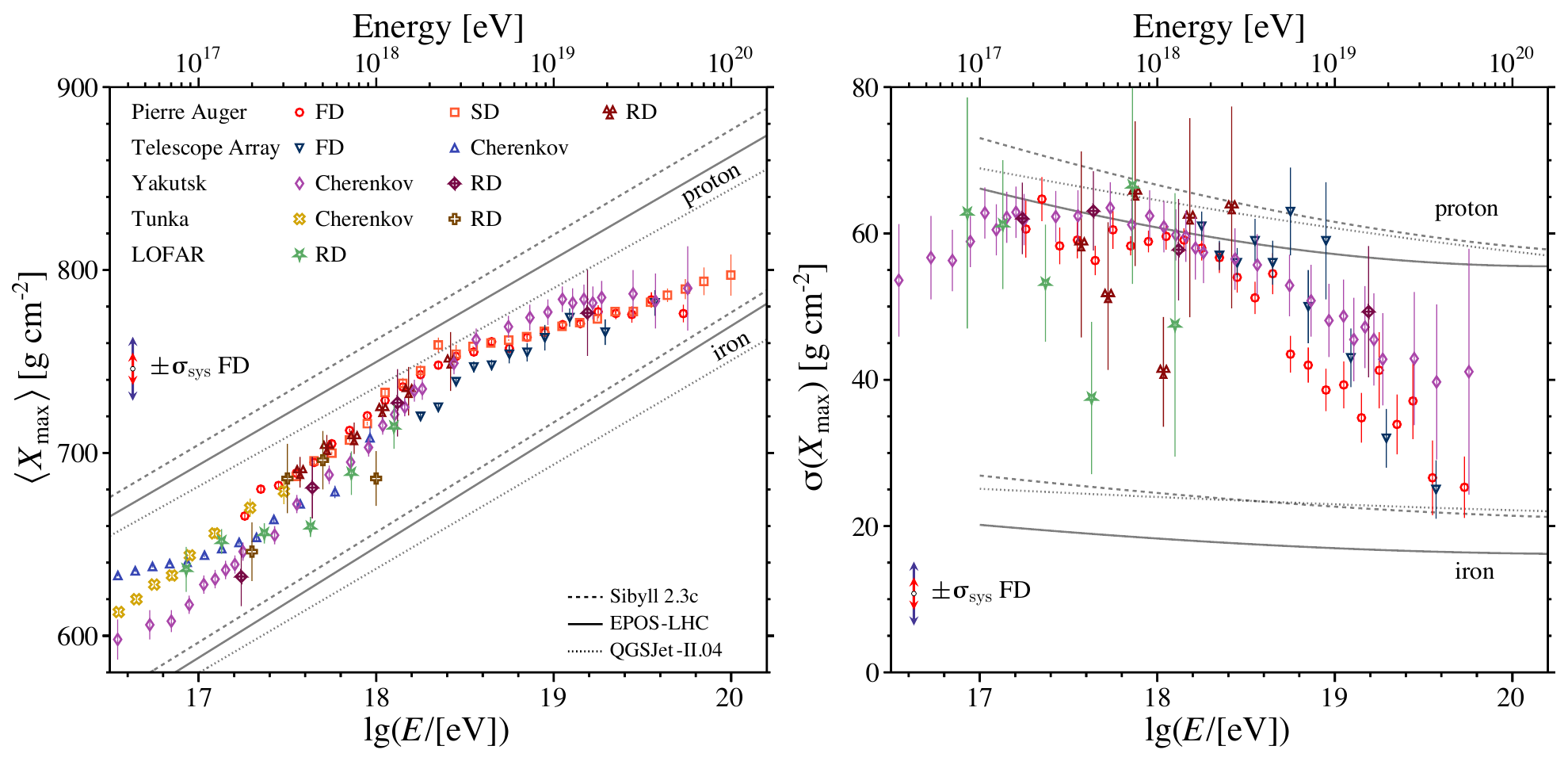}
\caption{In the left panel the mean of the distribution of the shower maximum and in the right panel its standard deviation as observed by Auger and TA. The TA data are corrected for detector effects by shifting $X_{max}$ by $+5$ g/cm$^2$ and by subtracting an $X_{max}$-resolution of 15 g/cm$^2$ (see \cite{Coleman:2022abf}). Figures taken from \cite{Coleman:2022abf}.}
\label{fig2}  
\end{figure}

In the left panel of figure \ref{fig2} we plot the mean $\langle X_{max} \rangle$ as measured by Auger and TA and in the right panel we plot the event-by-event fluctuations of the shower maximum distribution $\sigma(X_{max})$, the TA data are corrected subtracting detectors effects as discussed in \cite{Coleman:2022abf}. For comparison the prediction of $\langle X_{max} \rangle$ and $\sigma(X_{max})$ in the case of a pure proton and a pure iron composition are displayed. The expected behaviour of $\langle X_{max} \rangle$ and $\sigma(X_{max})$ differs depending on the hadronic interaction model used, in figure \ref{fig2} we plot three different models: EPOS-LHC (solid lines), Sybil 2.3 (dashed lines) and QGSJetII-04 (dotted lines) (see \cite{Coleman:2022abf,AlvesBatista:2019tlv} and references therein).

The determination of mass composition can also be pursued using the SD by measuring the number of muons of the EAS that reach the ground. This detection possibility is affected by larger uncertainties respect to the $X_{max}$ observation, due to theoretical unknowns in the hadronic physics at the highest energies of the EAS development. Experimentally, both Auger and TA observe a mismatch between the number of muons observed with energies above $10^{9.5}$ GeV and the number of muons predicted by the EAS models \cite{Coleman:2022abf,AlvesBatista:2019tlv}. 
 
On very general grounds we can state that the determination of the primary mass composition cannot be obtained in a model-independent way, the description of hadronic multi-particle production at the highest energies depends on the extrapolation of the accelerator data on cross sections up to center of mass energies $\sqrt{s}\simeq 400$ TeV. Moreover, the QCD predictions for the bulk of hadronic production at these energies are considerably affected by the hadronic interactions model assumed (see \cite{Coleman:2022abf,AlvesBatista:2019tlv} and references therein).

The most relevant signals of UHECR anisotropies are those connected with large scales. Auger observed, with a remarkable statistical significance, a dipole anisotropy at energies above $8\times 10^{18}$ eV, with and amplitude at the level of $6\%$ and the phase pointing toward the galactic anti-center, signalling a clear extragalactic origin of these particles \cite{Coleman:2022abf,AlvesBatista:2019tlv}. At lower energies, down to $10^{18}$ eV, the dipole amplitude has only upper limits at the level of $1\%$, with a phase that rotates with decreasing energy reaching the galactic center direction below $10^{18}$ eV, where the transition between galactic and extragalactic cosmic rays seems placed \cite{Coleman:2022abf,AlvesBatista:2019tlv,Aloisio:2017ooo,Aloisio:2017qoo}. 

Studies of small scale anisotropies aimed at identifying sources are not conclusive, with hints of clustering not yet at the required statistical significance (see \cite{Coleman:2022abf,AlvesBatista:2019tlv} and references therein).

\section{Sources of UHECR}
\label{sec:acc}
To constrain the basic features of the sources of UHECR it is useful to adopt a purely phenomenological approach in which sources are homogeneously and isotropically distributed, characterised by few parameters: the injection of particles by the sources with a power law behaviour in energy $\propto E^{-\gamma_g}$, the maximum acceleration energy $E_{max}$ and the source emissivity ${\cal L}_S(A)$ , i.e. number of particles injected per unit volume and energy for each nuclei species labeled by the atomic mass number $A$. These parameters, that univocally identify a class of sources, are fitted to experimental data of both spectrum and mass composition (see \cite{Coleman:2022abf,AlvesBatista:2019tlv,Aloisio:2017ooo,Aloisio:2017qoo} and references therein). 

\begin{figure}[!h]
\centering
\includegraphics[scale=.28]{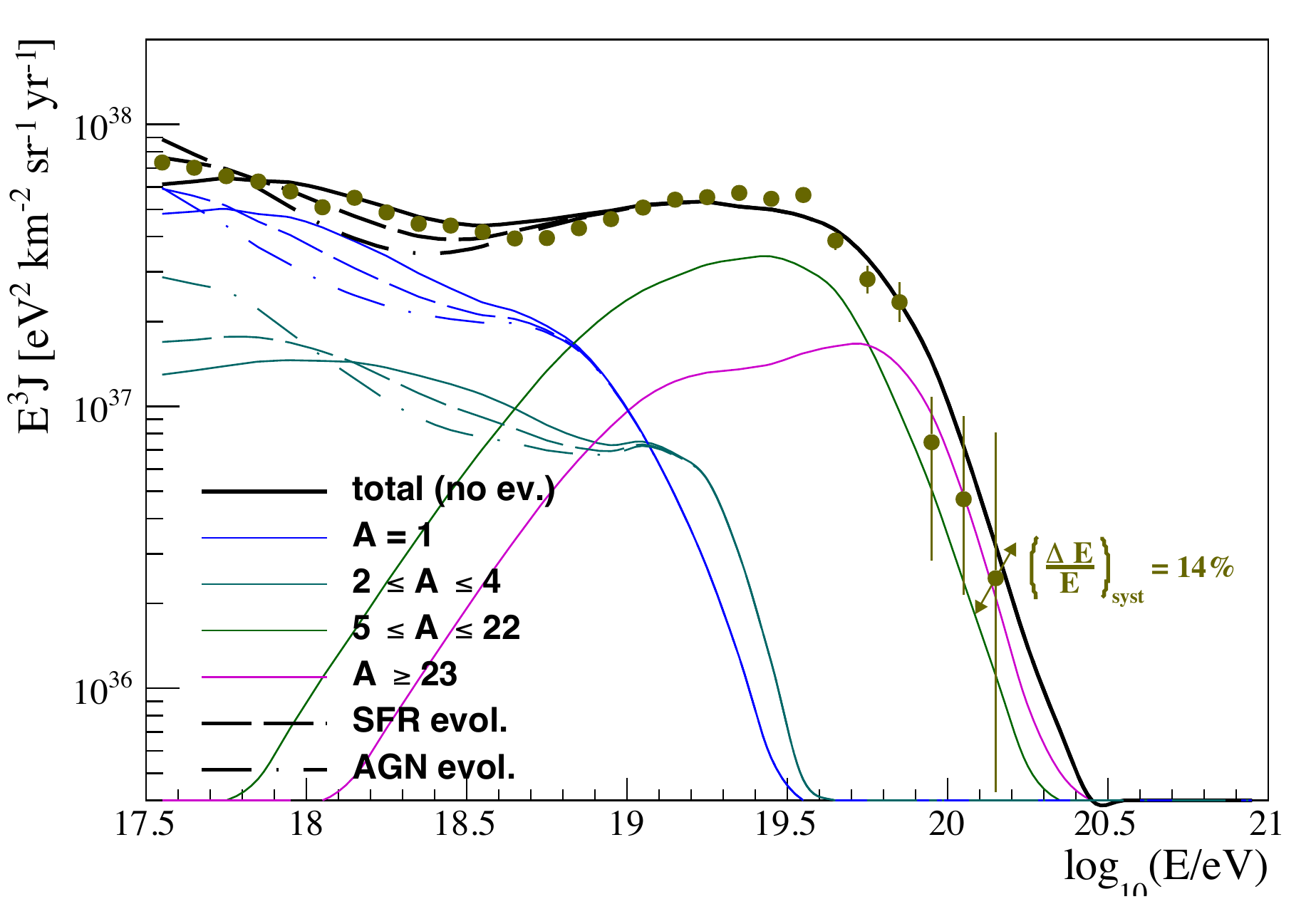}
\includegraphics[scale=.28]{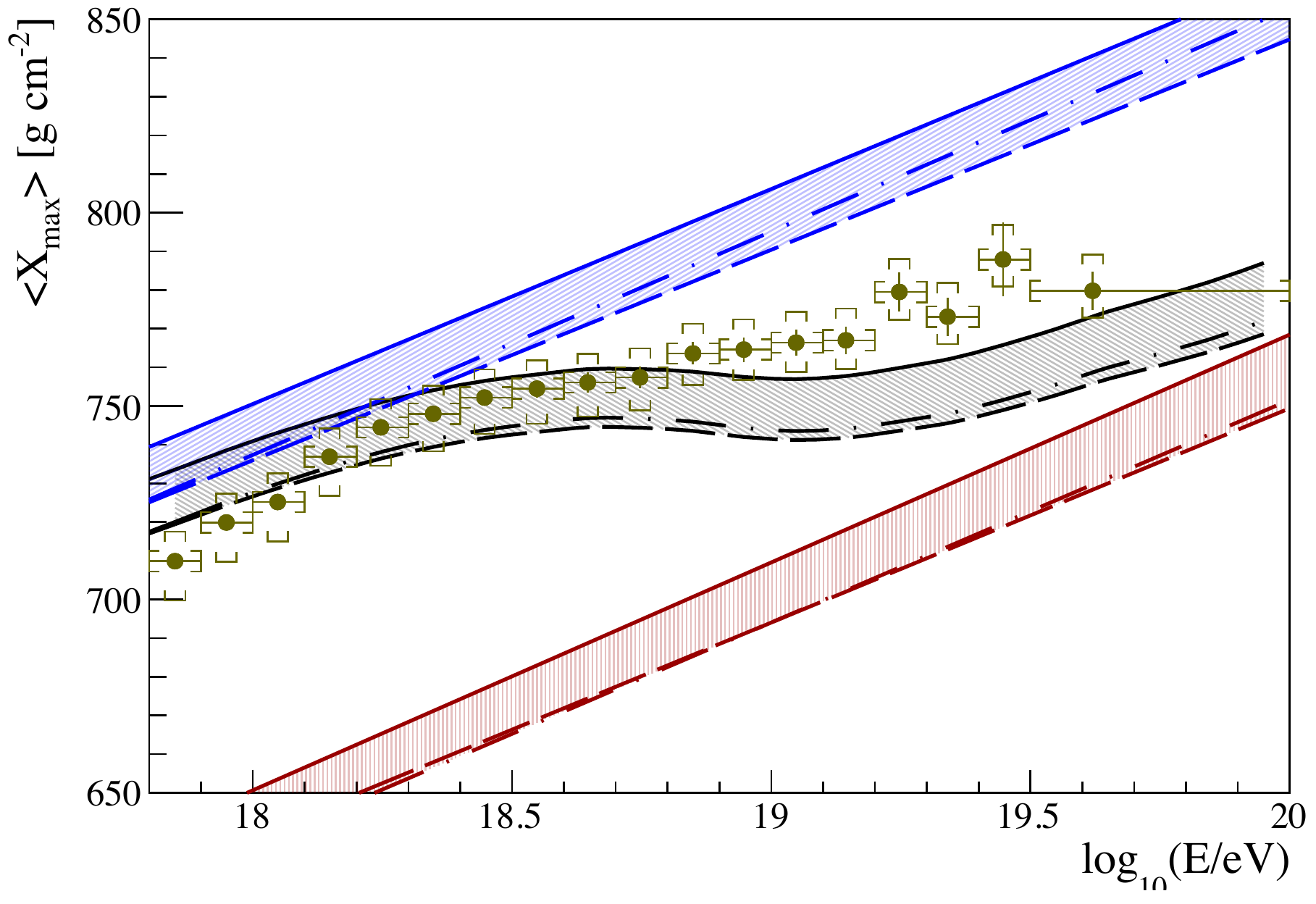}
\includegraphics[scale=.28]{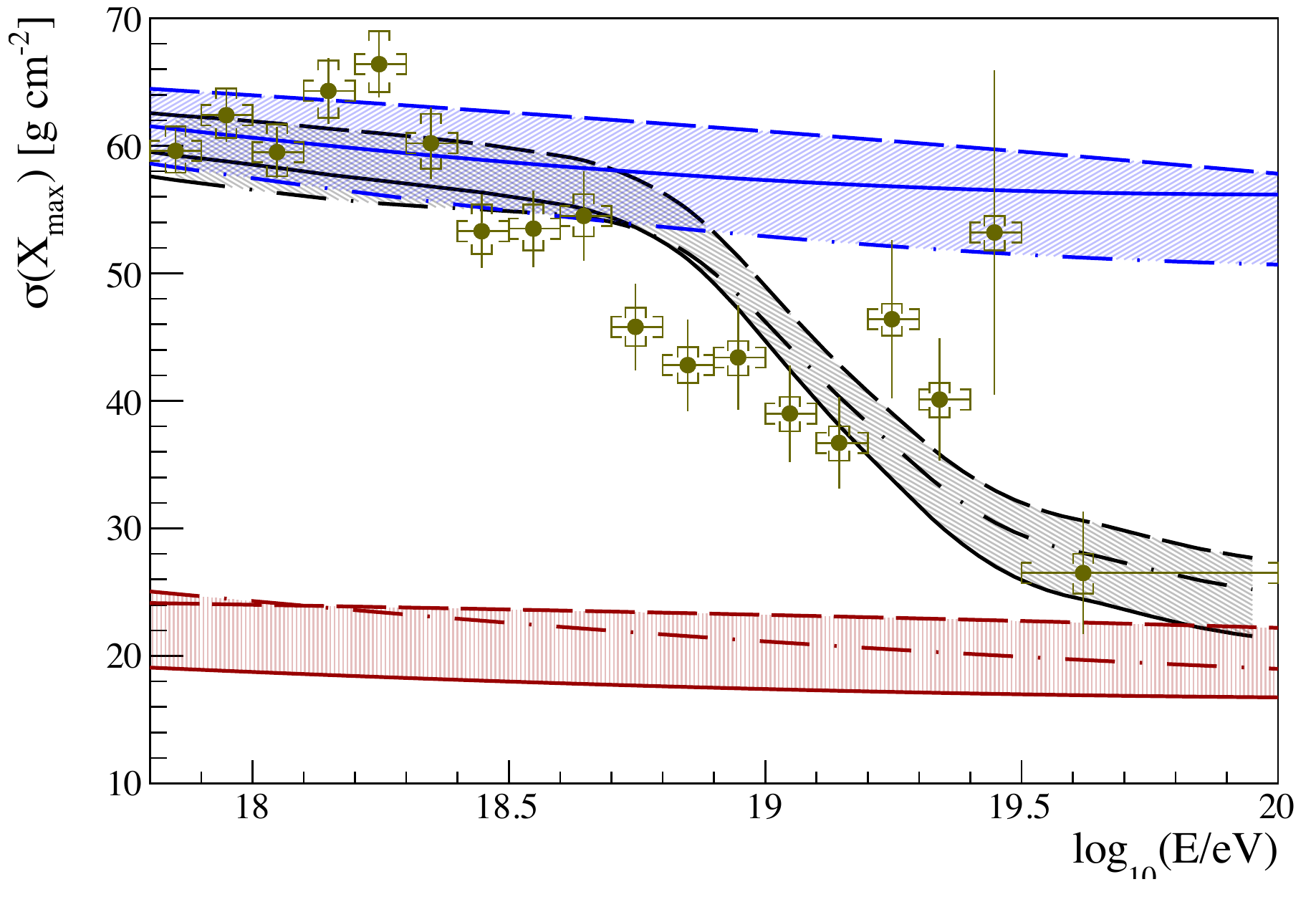}
\caption{Comparison of the flux (left panel), elongation rate (central panel) and its dispersion (right panel) as observed by Auger and computed assuming the model with two classes of different extragalactic sources (see text). Coloured bands show the uncertainties related to the hadronic interaction model. Figures taken from \cite{Aloisio:2015ega}.}
\label{fig3}  
\end{figure}

Following \cite{Aloisio:2013hya,Aloisio:2015ega}, taking into account all possible channels of energy losses and solving the transport equations for UHECR (protons or nuclei) we can determine the theoretical flux and mass composition expected at the Earth, to be compared with observations. While the flux directly follows solving the transport equations, mass composition is inferred from the mean value of the depth of shower maximum (elongation rate) $\langle X_{max} \rangle$ and its dispersion $\sigma(X_{max})$, computed as shown in \cite{PierreAuger:2013xim}. Once the TA data are corrected for the detector effects, as follows from figure \ref{fig2}, there is a substantial agreement between the results on mass composition of Auger and TA. Therefore, in the following discussion we will refer only to Auger data. 

The qualitative new finding that mass composition might be mixed has served as a stimulus to build models that can potentially explain the phenomenology of Auger data. These models all show that the Auger spectrum and mass composition at $E\ge 3\times 10^{18}$ eV can be fitted at the same time only at the price of requiring very hard injection spectra for all nuclei and a maximum acceleration energy $E_{max}\le 5 Z\times 10^{18}$ eV \cite{Aloisio:2013hya,Aloisio:2015ega,Aloisio:2009sj}, being $Z$ the charge of the nucleus. The need for hard spectra can be understood taking into account that the low energy tail of the flux of UHECR reproduces the injection power law. Therefore, taking $\gamma\ge 2$ cause the low energy part of the spectrum to be polluted by heavy nuclei thereby producing a disagreement with the light composition observed at low energy.

One should appreciate the change of paradigm that these findings imply: while in the case of a pure proton composition it is needed to find sources and acceleration mechanisms able to energise CR protons up to energies larger than $10^{20}$ eV with steep injection \cite{Aloisio:2006wv} ($\gamma_g\simeq 2.5\div 2.7$), the Auger data require that the high energy part of the spectrum ($E>3\times 10^{18}$ eV) has a flat injection ($\gamma_g\simeq 1.0\div 1.6$) being dominated by heavy nuclei with maximum energy not exceeding a few$\times Z\times 10^{18}$ eV  \cite{Aloisio:2013hya}.  In order to reproduce Auger observations, the additional light contribution to the flux at energies below $3\times 10^{18}$ eV should exhibit a steep power law injection with $\gamma_g\simeq 2.6\div 2.7$ and a maximum acceleration energy not exceeding a few$\times 10^{18}$ eV, as for the heavier component \cite{Aloisio:2013hya,Taylor:2013gga,Globus:2015xga}.

The origin of UHECR can be modelled essentially in two ways: (i) assuming the presence of different classes of sources: one injecting heavy nuclei with hard spectrum and the other only proton and helium nuclei with soft spectrum \cite{Aloisio:2013hya,Taylor:2013gga} or (ii) identifying a peculiar class of sources that could provide at the same time a steep light component and a flat heavy one \cite{Globus:2015xga,Unger:2015laa,Blasi:2015esa}. The second approach is based on a specific hypothesis on the sources that should be surrounded by an intense radiation field that, through photo-disintegration of heavy nuclei in the source neighbourhood, can provide a light component of (secondary) protons with a steep spectrum together with a hard and heavier component \cite{Globus:2015xga,Unger:2015laa}. It should be also noted that the diffusive shock acceleration mechanism produces steep injections of particles with $\gamma_g > 2$, while acceleration due to potential drops typically imply $\gamma_g\simeq 1$ (see \cite{Aloisio:2017ooo,Aloisio:2017qoo} and references therein). 

In figure \ref{fig3} we plot the flux and mass composition as observed by Auger and as reproduced theoretically assuming two classes of extragalactic sources with different injection characteristics as discussed above.

\section{Super Heavy Dark Matter}
\label{sec:shdm}

The leading paradigm to explain DM observations is based on the Weakly Interactive Massive Particle (WIMP) hypotheses, a stable DM particle with mass in the range of $0.1\div 10$ TeV \cite{Bergstrom:2000pn}, as it follows in the framework of super-symmetry solutions of the "naturalness problem". Searches for WIMPs are ongoing through three different routes: direct detection, indirect detection, and accelerator searches. None of these efforts have discovered a clear WIMP candidate so far. In addition, no evidence for new physics in the $0.1\div 10$ TeV range has been observed at the Large Hadron Collider (LHC). Although not yet conclusive, the lack of evidence for WIMP may imply a different solution for the DM problem outside of the WIMP paradigm.

To solve the DM puzzle, an alternative to WIMP models  is represented by the scenarios based on long lived super-heavy relics, that can be produced by several mechanisms taking place during the inflationary phase or just after, in the re-heating phase (see \cite{PierreAuger:2022ibr,PierreAuger:2022wzk,Aloisio:2006yi,Aloisio:2015lva,Aloisio:2007bh} and references therein). Once created in the early universe, the abundance of the long-lived super-heavy particles can evolve to match the DM density observed today, the so-called Super Heavy Dark Matter (SHDM). This conclusion can be drawn under three general hypotheses: (i) SHDM in the early universe never reaches local thermal equilibrium; (ii) SHDM particles have mass $M_X$ of the order of the inflaton mass or higher; and (iii) SHDM particles are long-living particles with a lifetime exceeding the age of the universe, $\tau_X\gg t_0$. 

Recently, using the LHC measurements of the masses of the Higgs boson and Top quark, has been pointed out that the Higgs potential is stable until very high energies \cite{Buttazzo:2013uya,Alekhin:2012py,Bednyakov:2015sca}. The scale $\Lambda_I$ at which the quartic coupling $\lambda$ of the Higgs potential becomes negative at large field values turns out to be $\Lambda_I\simeq 10^{10} \div 10^{12}$ GeV, moreover, being the $\lambda$ running extremely slow, this instability develops slowly and it might be possible to extrapolate the SM to even higher energies up to the Plank mass $M_P=10^{19}$ GeV, with no need for new physics until ultra high energies \cite{Buttazzo:2013uya}. This fact implies that the mass spectrum of the Dark Sector (DS) is restricted to ultra high energies and SHDM could play a role. In the following we will consider SHDM with masses in the range between $10^{8}$ GeV up to the Plank mass. 

Apart from the gravitational interaction, SHDM can be coupled to ordinary matter also through some super-weak coupling, driven by a high energy scale $\Lambda>\Lambda_I$ as the Grand Unification Scale ($\Lambda_{GUT} \simeq 10^{16}$ GeV). To assure long-lived particles, the interaction SHDM-SM should be suppressed by some power $n$ of the high energy scale $\Lambda$, with a decay SHMD $\rightarrow$ SM characterised by a lifetime $\tau_X\simeq (M_X \alpha_{X\Theta})^{-1} (\Lambda/M_X)^{2n-8}$, being $\alpha_{X\Theta}$ the reduced coupling constant between SHDM and SM particles (see \cite{PierreAuger:2022ibr} and references therein). 

In the case in which SHDM interacts with SM particles only through the gravitational interaction, assuming the very general case of a DS characterised by its own non-abelian gauge symmetry, SHDM can decay only through non-perturbative effects as the instanton-induced decay (see \cite{PierreAuger:2022ibr} and references therein). In this case the lifetime of SHDM follows from the corresponding instanton transition amplitude that, being exponentially suppressed, provides long-living particles. Considering the zeroth oder contribution, the instanton-induced lifetime can be written as $\tau_X\simeq M_X^{-1} \exp{(4\pi/\alpha_X)}$, being $\alpha_X$ the reduced coupling constant of the hidden gauge interaction in the DS.   

\begin{figure}[!h]
\centering
\includegraphics[scale=.265]{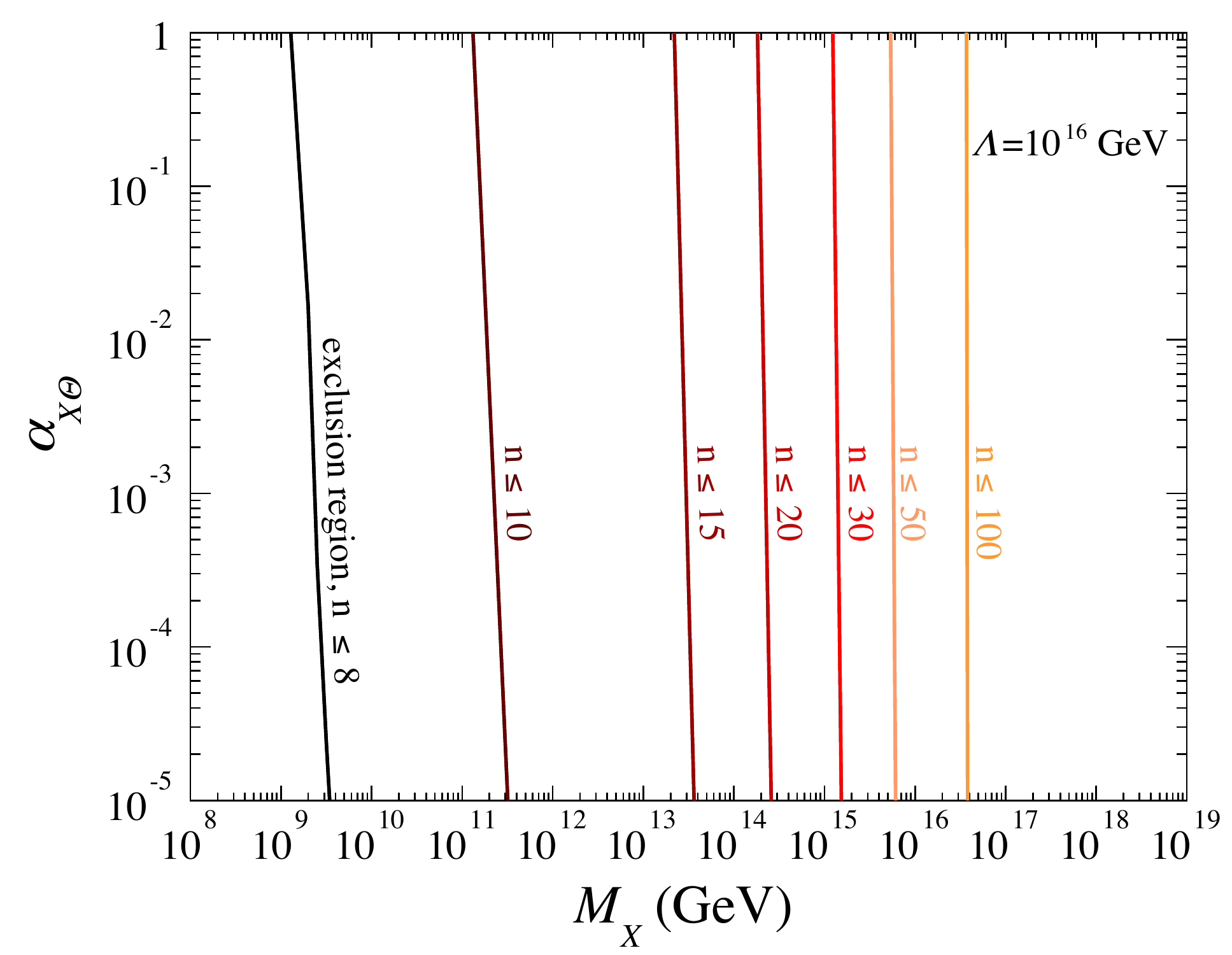}
\includegraphics[scale=.25]{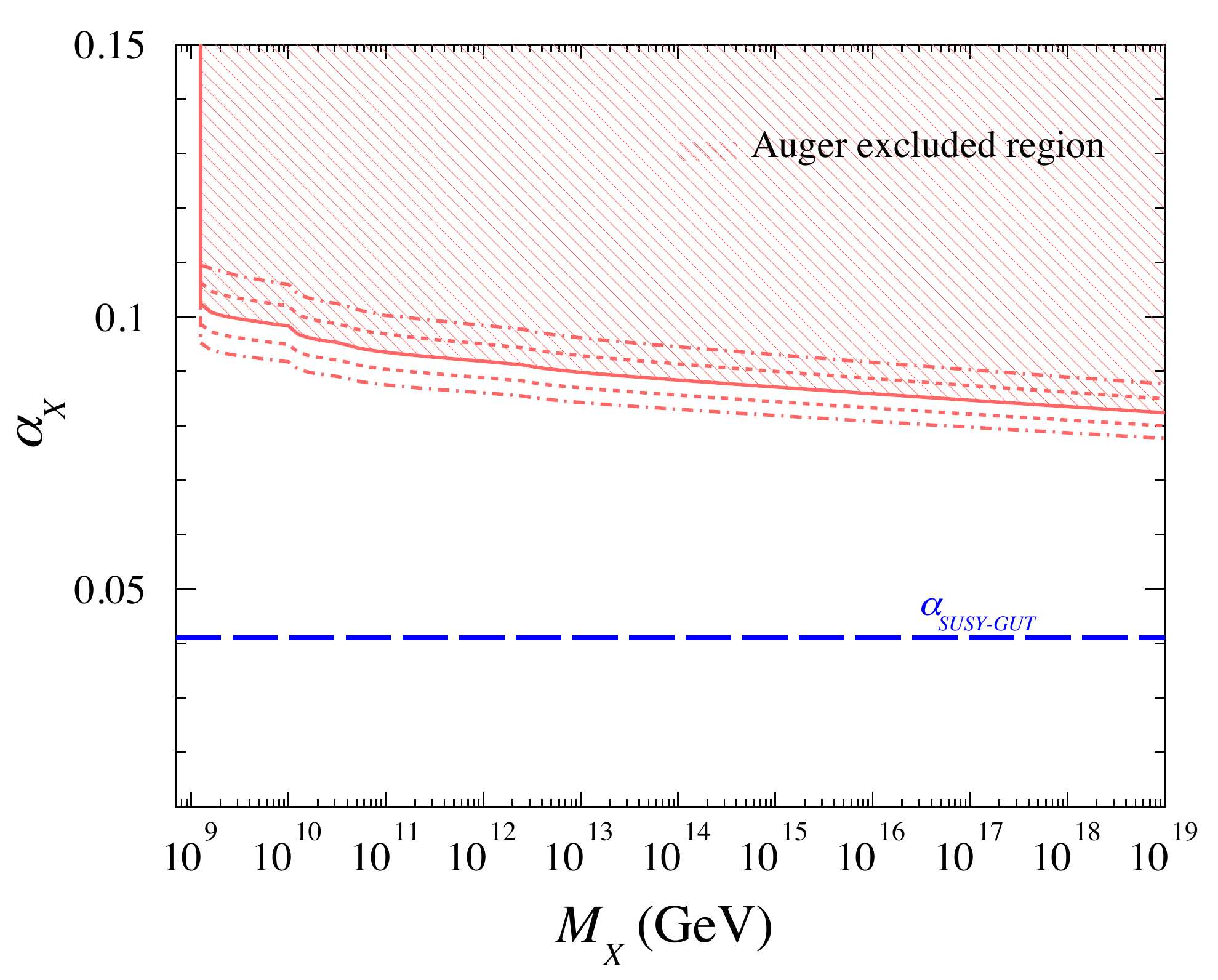}
\includegraphics[scale=.25]{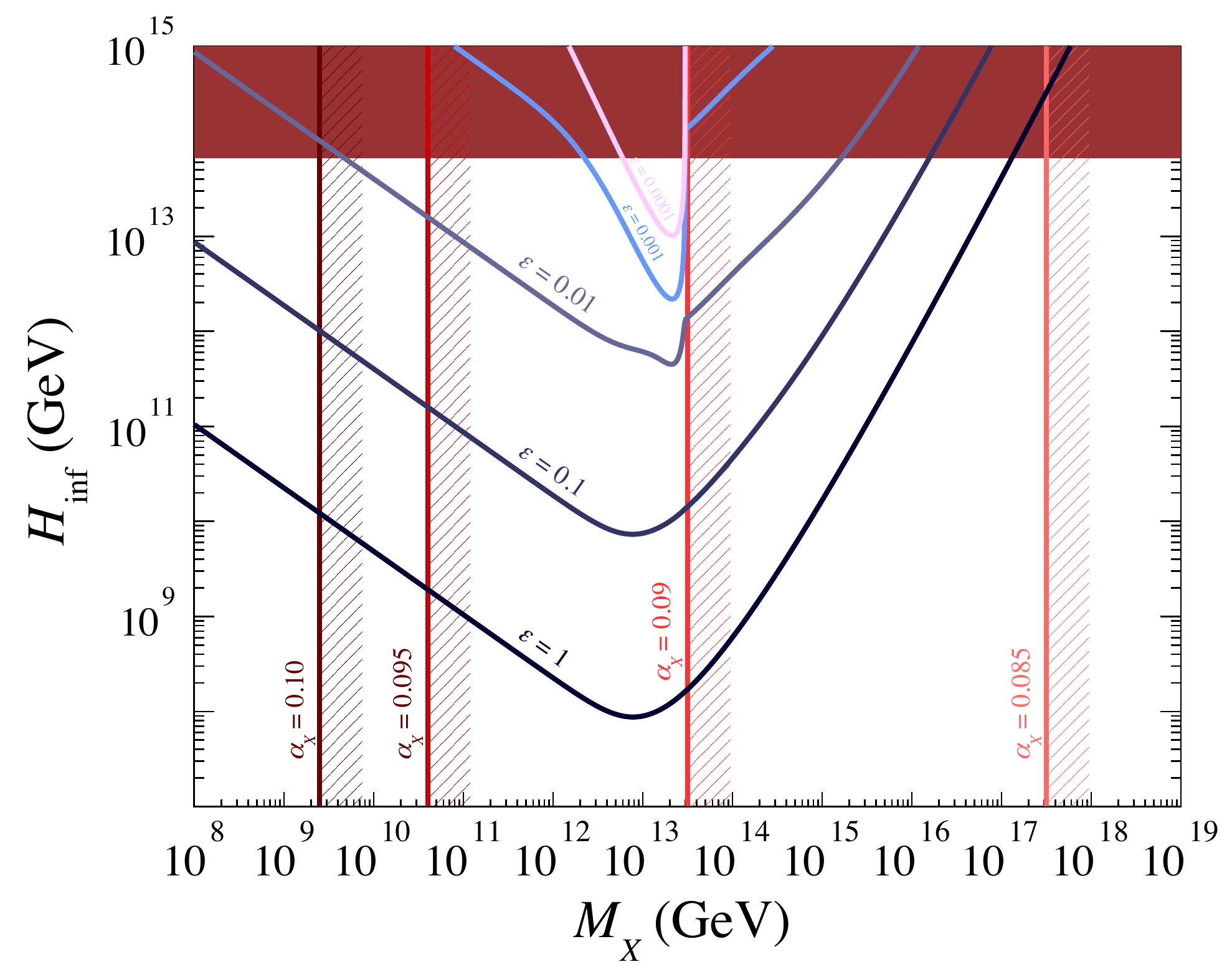}
\caption{Limits fixed by the Auger observations on the SHDM mass $M_X$ and: (left panel) the coupling SHDM-SM $\alpha_{X\Theta}$ for different values of the suppression scale order $n$, (central panel) the coupling of the hidden DS gauge interaction $\alpha_{X}$ in the case of the instanton-induced decay, (right panel) the universe expansion rate at the end of inflation $H_{inf}$ for different values of the reheating efficiency $\epsilon$ (see text). Figures taken from \cite{PierreAuger:2022ibr}.}
\label{fig4}  
\end{figure}

Under very general assumptions on the underlying theory (see \cite{Adhikari:2022sve,PierreAuger:2022ibr,PierreAuger:2022wzk,Aloisio:2006yi,Guepin:2021ljb,Aloisio:2015lva,Aloisio:2007bh} and references therein), we can determine the composition and spectra of the standard model particles produced by the SHDM decay. Typical decay products are couples of quark and anti-quark that, through a cascading process (jets), give rise to neutrinos, gamma rays and nucleons with a flat spectrum, that at the relevant energies can be approximated as $dN/dE \propto E^{-1.9}$, independently of the particle type, with a photon/nucleon ratio of about $\gamma/N\simeq 2\div 3$ and a neutrino nucleon ratio $\nu/N\simeq 3\div 4$, quite independent of the energy range \cite{Aloisio:2006yi}. The most constraining limits on the SHDM mass and lifetime are those coming from the (non) observation of UHE photons and, even to a lesser extent, neutrinos in the UHECR regime. Auger observations provide us with very stringent limits on the photon and neutrino fluxes in the energy range $10^{8}\div 10^{11}$ GeV, enabling far the best limits on SHDM models \cite{PierreAuger:2022ibr,PierreAuger:2022wzk}. 

In the left and central panels of figure \ref{fig4} we plot the limits fixed by the Auger observations on the SHDM mass $M_X$ and the coupling SHDM-SM $\alpha_{X\Theta}$, in the case of direct coupling (left panel), and the coupling of the hidden DS gauge interaction $\alpha_{X}$ in the case of the instanton-induced decay (central panel). Requiring that the SHDM density today fits the observed DM density, the limits on $M_X$ and $\tau_X$ can be rewritten in terms of the cosmological parameters characterising the generation mechanism of SHDM. In the right panel of figure \ref{fig4} we plot the limits coming from Auger data on the age of the universe at the end of inflation $H_{inf}^{-1}$ and the efficiency of the reheating phase $\epsilon=(\Gamma_\phi/H_{inf})^{1/2}$, being $\Gamma_\phi=g_\phi^2 M_\phi/8\pi$ the inflaton (of mass $M_\phi$) decaying amplitude (see \cite{PierreAuger:2022ibr,PierreAuger:2022wzk} and references therein). 

\section{Conclusions}
\label{sec:concl}

Thanks to the measurements of Auger and TA, in the past 20 years the physics of UHECR experienced a paradigm shift. The simple picture of protons at the highest energies has been replaced by a more complex and (phenomenologically) richer one with nuclei dominating the spectrum already at energies around $5\times 10^{18}$ eV. Moreover, at the highest energies, it was established the existence of a strong suppression in the spectrum of UHECR. Despite these important achievements some long-standing problems in the physics of UHECR remain unsolved: what are the sources and which astrophysical mechanism accelerate particles? What is the mass composition at the highest energies? What is the flux of neutrinos and gamma rays produced by UHECR interactions at the source and during propagation?

In the next decade the upgrades of the Auger and TA observatories can contribute answering these questions \cite{Coleman:2022abf,AlvesBatista:2019tlv}. Yet, to reach this goal, it is necessary to refine both experimental analysis and theoretical understandings. Mass determination is currently limited by the uncertainties in the predictions of hadronic models and muon content in the EAS, the planned Auger upgrade with an independent measurement of the electromagnetic and muonic content of the EAS could help solve this problem. Precise mass determination till the highest energies will be of paramount importance in the future. The necessary step forward should be disentangling the all-particle energy spectrum into that of individual mass groups (as for instance p, He, CNO, MgAlSi, Fe). On more general grounds, the clearest path to event-by-event primary mass reconstruction lies in a high-resolution independent reconstruction of both $X_{max}$ and muon content coupled to a high-resolution energy measurement.

The behaviour of the spectrum at the highest energies is not univocally determined, as the results of Auger and TA, even in the same declination band, seem not compatible, with an earlier position of the flux suppression energy in the Auger dataset. In order to clarify this point it would be very useful if the two collaborations could use the same fluorescence yield and invisible energy to analyse the data. Moreover, the origin of the flux suppression at the highest energies is still uncertain. Solving the puzzle between a propagation generated suppression or an acceleration generated one requires individual mass groups energy spectra and a detailed theoretical modelling of particles acceleration and propagation in the source environment. 

As follows from the observed mass composition a clear identification of individual sources would be difficult. This goal requires a thorough theoretical modelling of the effects of galactic and extragalactic magnetic fields on UHECR propagation and a shower-by-shower determination of the mass and energy of the primary particle. 
The study of UHECR has an impact also in fundamental physics, as it involves tests of models and theories that extend beyond the standard model of particle physics. This is the case of SHDM that, being a viable alternative to the WIMP paradigm for DM, can be tested only through the observations of UHECR eventually connected with those of cosmology \cite{Adhikari:2022sve,Aloisio:2015lva}, as follows from the right panel of figure \ref{fig4}. 

We conclude stating that a complete picture of the ultra high energy universe needs to account for all messengers till the highest energies. The sensitivity required to detect UHE neutrinos and gamma rays, as well as the highest energy tail of CR, is still far from being reached, new technologies are needed and future space observatories, with improved photon detection techniques, allowing the needed statistics, might open a new era in the physics of UHE cosmic radiation \cite{Cummings:2020ycz,Cummings:2021bhg,POEMMA:2020ykm}.
 

\end{document}